\documentclass[aps,pra,twocolumn,floats,amsmath,amssymb,superscriptaddress]{revtex4-1}
\usepackage{graphicx}
\usepackage{epsfig}
\usepackage{amsfonts}
\usepackage{amsmath}
\usepackage{natbib}
\usepackage{multirow}
\usepackage{bm}
\usepackage{epstopdf}
\usepackage{ulem}
\usepackage{color}
\usepackage{hyperref}
\usepackage{cleveref}
\usepackage{ marvosym }
\usepackage{rotating}
\usepackage{ tipa }

\begin{document}
\title{Structural, electronic, and optical properties of periodic graphene/h-BN van der Waals heterostructures}
\author{Wahib Aggoune} 
\email{Corresponding author: wahib.aggoune@physik.hu-berlin.de}
\affiliation{Institut f\"{u}r Physik and IRIS Adlershof, Humboldt-Universit\"{a}t zu Berlin, 12489 Berlin, Germany}
\affiliation{Laboratoire de Physique Th\'eorique, Facult\'e des Sciences Exactes, Universit\'e de Bejaia, 06000 Bejaia, Algeria}

\author{Caterina Cocchi}
\email{Present address: Institut f\"{u}r Physik, Carl von Ossietzky Universit\"{a}t, Carl-von-Ossietzky-Stra\ss e 9, 26129 Oldenburg, Germany.}
\author{Dmitrii Nabok}
\affiliation{Institut f\"{u}r Physik and IRIS Adlershof, Humboldt-Universit\"{a}t zu Berlin, 12489 Berlin, Germany}
\affiliation{European Theoretical Spectroscopic Facility (ETSF)}

\author{Karim Rezouali}
\author{Mohamed Akli Belkhir}
\email{Present address: Present address: Universit\'e Batna 2, 53 Route de Constantine, Fesdis, Batna 05078, Alg\'erie.}
\affiliation{Laboratoire de Physique Th\'eorique, Facult\'e des Sciences Exactes, Universit\'e de Bejaia, 06000 Bejaia, Algeria}

\author{Claudia Draxl}
\affiliation{Institut f\"{u}r Physik and IRIS Adlershof, Humboldt-Universit\"{a}t zu Berlin, 12489 Berlin, Germany}
\affiliation{European Theoretical Spectroscopic Facility (ETSF)}

\begin{abstract}
The emerging interest in van der Waals heterostructures as new materials for opto-electronics and photonics poses questions about their stability and structure-property relations. In the framework of density-functional  and many-body perturbation theory, we investigate the structural, electronic, and optical properties of periodic heterostructures formed by graphene and hexagonal boron nitride (h-BN). To understand how the constituents affect each other depending on the layer stacking, we examine 12 commensurate arrangements. We find that interaction with graphene improves the stability of bulk h-BN also in those configurations that are predicted to be energetically metastable. In return, the interaction with h-BN can open a band gap of a few hundred meV in graphene. Its actual size can be tuned by the arrangement of the layers. In the semiconducting configurations, the character and spatial distribution of optical excitations are affected by the specific stacking, that determines the electronic states involved in the transitions. Remarkably, six out of the 12 explored heterostructures remain semi-metallic. 
\end{abstract}
\maketitle

The possibility of producing and engineering low-dimensional systems at the nanoscale has
broadened the horizon of materials science. Among the most promising candidates for a new
generation of opto-electronic and photonic devices, graphene and other monolayer materials play a special role,
due to their unique and tunable electronic properties~\cite{geim+07nm,thyg17tdm}. Even more exciting perspectives have been opened by combining these systems in heterostructures, thereby exploiting and enhancing the characteristics of the single components~\cite{geim-grig13nat,novo+16sci}. Recently, van der Waals heterostructures (vdWh) formed by graphene and hexagonal boron nitride (h-BN) have received extensive attention~\cite{haigh+12nm,brit+12sci,britnel+13sci,wood+14natp,chen+14natcom,mish+14natn,li+16small,saka+15jpsj,pierucci+apl18,yanko+19nrp}. This combination is motivated by the relatively small lattice mismatch ($<$1.7~$\%$) between the constituents and by the possibility to open a band gap in graphene~\cite{rama+11nl,Hunt+13sc,giov+07prb,khar-saro11nl,quhe+12npgam,ritwika+15mre}. Such heterostructure has been reported to be an excellent platform for graphene-based plasmonic devices~\cite{woessner+14nm, alessandro+14prb}. Moreover, the optical conductivity of graphene~\cite{stauber+08prb} has been found to be tunable in the visible range by rotating the graphene sheet on the h-BN substrate~\cite{stotman+15prl}. Finally, in a previous work~\cite{agg+17jpcl}, we have shown that, when intercalating graphene layers with h-BN in a periodic vdWh, the intrinsic optical characteristics of the constituents are to a large extent preserved, but furthermore, exhibit peculiar optical excitations that are not present in the individual building blocks~\cite{agg+17jpcl}. In particular, the stacking has a strong impact on the spatial distribution of the resulting electron-hole (e-h) pairs~\cite{agg+17jpcl}. 

In real samples, the deposition of graphene on h-BN leads to the formation of Moir$\acute{\mathrm{e}}$ patterns due to the small in-plane lattice mismatch~\cite{wood+14natp,decher+11nl,wijk+14prl,argentero+17nl,yankwitz+12np}. In this case, the stacking is inhomogeneous, and the layers corrugate in the out-of-plane direction, giving rise to locally different interlayer distances~\cite{argentero+17nl}. It was reported that the twist angle, the lateral displacement, and the degree of strain promote local transitions from incommensurate to commensurate arrangements between graphene and h-BN~\cite{decher+11nl,kim+17nl,sanjose+14prb,wood+14natp}, such that homogeneous stackings can be formed. The variety of structural arrangements in terms of interlayer distances, atomic displacements, and layer stackings can be exploited for fine-tuning the electronic and optical properties of graphene/h-BN heterostructures. 

To rationalize the behavior of such complex systems, we explore here commensurate graphene/h-BN vdWh, focusing on how the intrinsic electronic and optical properties are modified upon stacking. To this end, we consider all inequivalent configurations allowed in the hexagonal unit cells, where no lattice rotations are included. These structures are constructed by sandwiching a graphene monolayer between two h-BN layers and applying periodic boundary conditions. We employ density functional theory (DFT) and many-body perturbation theory (MBPT) to compute groundstate and excited-state properties. First, we show how the structural stability of vdWh can be enhanced by layer stacking. Then, we analyze their electronic properties, discussing under which conditions the h-BN layers may or may not open a band gap in graphene and how, in return, graphene leads to a band-gap renormalization of h-BN. We finally explore the optical excitations in the semiconducting vdWh, focusing on how their character and spatial distribution are affected by stacking. 
\section*{Theoretical background and computational details}
Ground-state properties are calculated using DFT~\cite{hohe-kohn64pr,kohn-sham65pr} with the generalized gradient approximation in the Perdew-Burke-Ernzerhof (PBE) parameterization \cite{PBE} for the exchange-correlation functional. To account for van der Waals interactions between the layers, we adopt the semi-empirical DFT-D2 method~\cite{grim06jcc}. For ranking the stability of the explored stackings, we evaluate the formation energy as
\begin{equation}
E_{\mathrm{form}}= E_{\mathrm{tot}}^{\mathrm{vdWh}}-[ E_{\mathrm{tot}}^{\mathrm{g}}+2E_{\mathrm{tot}}^{\mathrm{h-BN}}],
\label{eq:stab}
\end{equation}
where the total energy of isolated graphene, $E_{\mathrm{tot}}^g$, and that of h-BN, $E_{\mathrm{tot}}^{\mathrm{h-BN}}$, are subtracted from the total energy of the heterostructure, $E_{\mathrm{tot}}^{\mathrm{vdWh}}$. The formation energies of graphene/h-BN and h-BN/h-BN bilayers are evaluated by subtracting the total energy of the single layers from the respective total energy of the bilayer.

Quasi-particle (QP) energies are computed within the $G_{0}W_{0}$ approximation \cite{hedi65pr,hybe-loui85prl} by solving the QP equation
\begin{equation}
\varepsilon_{i}^{QP}=\varepsilon_{i}
^{KS}+\langle\phi_{i}^{KS}\vert\Sigma(\varepsilon_{i}^{QP})
-v_{XC}^{KS}\vert\phi_{i}^{KS}\rangle,
\end{equation}
where $\Sigma$ is the non-local and energy dependent electronic self-energy, $\varepsilon_{i}^{KS}$ and $\phi_{i}^{KS}$ are the Kohn-Sham energies and wave-functions, respectively, and $v_{XC}^{KS}$ represents the exchange-correlation (xc) potential.

The optical spectra are obtained by solving the Bethe-Salpeter equation (BSE), the equation of motion of the two-particle Green function \cite{hank-sham80prb,stri88rnc}. This problem can be mapped onto the secular equation 
\begin{equation}
\sum_{v'c'\mathbf{k'}} H^{BSE}_{vc\mathbf{k},v'c'\mathbf{k'}}A^{\lambda}_{v'c'\mathbf{k'}} = E^{\lambda}A^{\lambda}_{vc\mathbf{k},}
\label{eq:ham}
\end{equation}
where $v$, $c$ and \textbf{k} indicate valence bands, conduction bands, and \textbf{k}-points in the reciprocal space, respectively. The effective Hamiltonian consists of three terms, $H^{BSE} = H^{diag} + H^{dir} + 2H^{x}$. The first term, $H^{diag}$, accounts for \textit{vertical} transitions between QP energies and, when considered alone, corresponds to the independent QP approximation (IQPA). The other two terms incorporate the screened Coulomb interaction ($H^{dir}$) and the bare electron-hole exchange ($H^{x}$). The factor 2 in front of the latter accounts for the spin multiplicity in non-spin-polarized systems. The eigenvalues of Eq.~\eqref{eq:ham}, $E^{\lambda}$, are the excitation energies. The corresponding eigenvectors, $A^{\lambda}_{vc\mathbf{k}}$, provide information about the composition of the $\lambda$-th excitation and act as weighting factors in the transition coefficients
\begin{equation}
\mathbf{t}_{\lambda} = \sum_{vc\mathbf{k}} A^{\lambda}_{v c \mathbf{k}} \frac{\langle v \mathbf{k} \vert \widehat{\mathbf{p}} \vert c \mathbf{k} \rangle}{\epsilon_{c \mathbf{k}}\ -\ \epsilon_{v \mathbf{k}}},
\label{eq:osci} 
\end{equation}
which determine the oscillator strength in the imaginary part of the macroscopic dielectric function,
\begin{equation}
\mathrm{Im}\varepsilon_M~=~\dfrac{8\pi^2}{\Omega} \sum_{\lambda} |\mathbf{t}_{\lambda}|^2 \delta(\omega - E^{\lambda}),
\label{eq:abs}
\end{equation}
where $\Omega$ is the unit cell volume.

All calculations are performed using \texttt{exciting}~\cite{gula+14jpcm,nabo+16prb,vor+es19}, an all-electron full-potential code, implementing the family of linearized augmented planewave plus local orbitals methods. Spin-orbit coupling (SOC) is not considered. In the ground-state calculations, a basis-set cutoff R$_{MT}$G$_{max}$=7 is used, where for all the atomic species involved, namely boron (B), nitrogen (N), and carbon (C), a muffin-tin radius R$_{MT}$ of 1.3~bohr is adopted. The sampling of the Brillouin zone (BZ) is carried out with a 30~$\times$~30~$\times$~8 $\textbf{k}$-grid. These parameters ensure a numerical precision of less than 1~meV in both the total energy and in the PBE band gap. Lattice constants and internal coordinates are optimized until the residual forces on each atom are less than 0.01~eV/\AA{}. In the $G_{0}W_{0}$ calculations~\cite{nabo+16prb}, 250 empty states are included to compute the frequency-dependent dielectric screening within the random-phase approximation. In this case, the BZ sampling is performed on a 18~$\times$~18 $\times$~4 shifted $\textbf{k}$-mesh for the semiconducting configurations. With these parameters, a numerical precision of about 40~meV is reached for the QP gap. In the case of the semi-metallic stackings, a 15~$\times$~15~$\times$~2 $\textbf{k}$-point mesh is adopted. For the solution of the BSE~\cite{vor+es19} on top of the QP electronic structure, a plane-wave cutoff R$_{MT}$G$_{max}$=6 is employed. The screened Coulomb potential is computed using 100 empty bands. 
In the construction and diagonalization of the BSE Hamiltonian, 3 occupied and 3 unoccupied bands are included and a 30~$\times$~30 $\times$~4 shifted $\textbf{k}$-point mesh is adopted to sample the BZ. Owing to the high dispersion around the high- symmetry point K, the \textbf{k}-point sampling is interpolated onto a 60~$\times$~60~$\times$~4 $\textbf{k}$-mesh, using the double-grid technique proposed in Refs.~\cite{gille+13prb,kamm+12prb}. This choice ensures converged spectra, specifically above 2~eV, and an accuracy on the binding energy of the lowest-energy exciton up to 10~meV.
Local-field effects are taken into account by including 41~$|\mathbf{G}+\mathbf{q}|$ vectors.
Atomic structures and isosurfaces are visualized using the VESTA software~\cite{momm-izum11jacr}.
\section*{Structural Properties}
\begin{figure*}
 \begin{center}
\includegraphics[width=.9\textwidth]{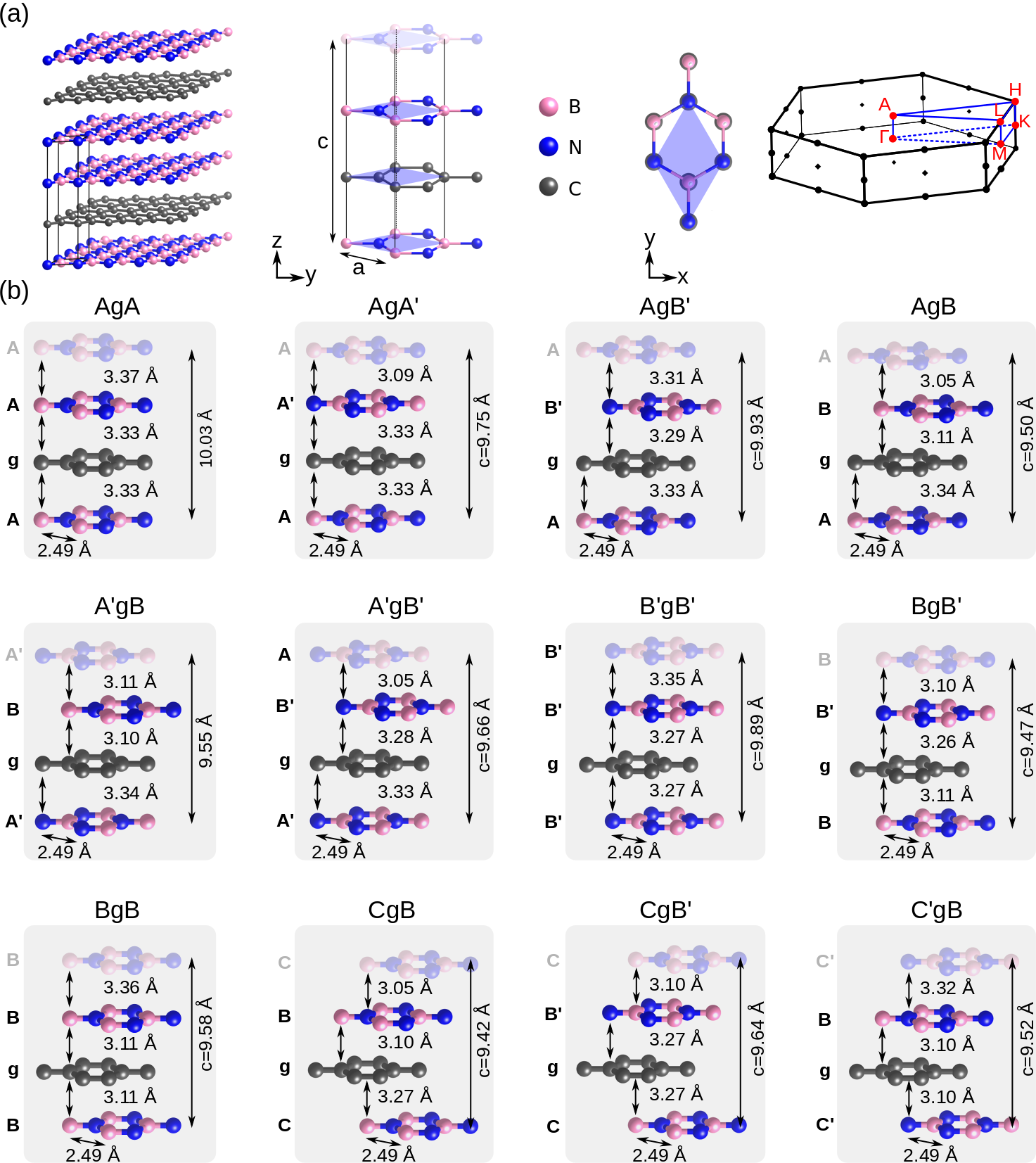}%
\caption{(a) From left to right: Sketch of the vdWh considered in this work; side and top views of the unit cell; first Brillouin zone with the high-symmetry points marked in red and the path connecting them in blue. (b) The twelve stackings considered in this work. The graphene sheet (\textit{g}) is taken as a reference while A, B, and C label the h-BN layers depending on their in-plane displacement with respect to graphene; the notation A', B', or C' is used when the positions of all boron and nitrogen atoms are swapped within the h-BN layer.}
\label{fig:hetero-unit cell}
 \end{center}
\end{figure*}
 The unit cells of all the graphene/h-BN vdWh considered in this work contain three layers with a total of six atoms (2 N, 2 B, and 2 C), where one graphene sheet is sandwiched between h-BN layers  [see Fig.~\ref{fig:hetero-unit cell}(a)]. In Fig.~\ref{fig:hetero-unit cell}(b) we show the twelve configurations examined in this work, also indicating their structural parameters. To simplify the nomenclature of the different stackings, we consider the graphene sheet (labeled (\rotatebox{15}{$g$})) as a reference within the unit cell. h-BN layer is
marked either by A, B, or C, depending on its relative in-plane displacement with respect to graphene. The notation A', B', or C' indicates the swapping of the position of all boron and nitrogen atoms within their reference A, B, or C layer, respectively.  As an example, in the A\rotatebox{15}{$g$}A and A\rotatebox{15}{$g$}A' configurations shown in Fig.~\ref{fig:hetero-unit cell}(b), all layers are aligned in the vertical direction, i.e. each B and N atom is sitting on top/below a C atom of graphene. While in the former B (N) atoms are on top of B (N) atoms, in the A\rotatebox{15}{$g$}A' configuration, B and N atoms lie on top of each other in adjacent h-BN layers. By shifting one h-BN layer by one bond length in the in-plane direction, we obtain the A\rotatebox{15}{$g$}B' stacking [see Fig.~\ref{fig:hetero-unit cell}(b)]. 
\begin{figure*}[htb]
 \begin{center}
\includegraphics[width=.9\textwidth]{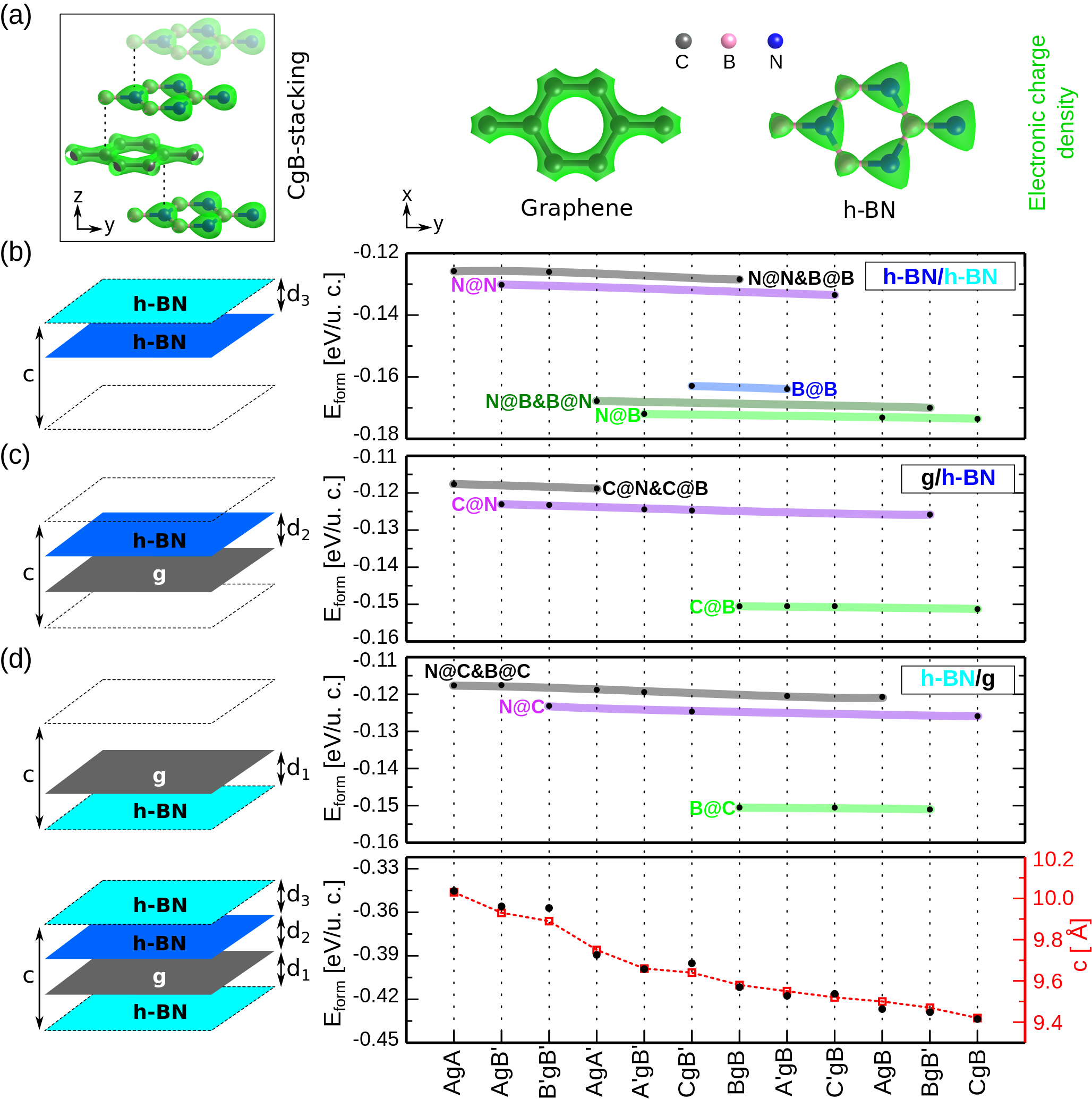}%
\caption{(a) Electronic charge density in the C
$\textscriptg$B heterostructure; left:  side view, right: top view of graphene and h-BN layers. Formation energy computed for b) the two inequivalent h-BN layers, c)-d) graphene and the h-BN layer above and below it. The symbol @ indicates vertically aligned atoms. (Bottom panel) Out-of-plane lattice parameter \textit{c} and formation energy of the full heterostructures.}
\label{fig:stability}
 \end{center}
\end{figure*}
The calculated in-plane lattice parameters \textit{a}=2.49~\AA{} of the vdWh remains the same in all the considered configurations, as it is determined by covalent bonding within the planes. The corresponding calculated in-plane lattice parameters of pristine graphene and bulk h-BN~\cite{agg+18prb} are 2.46~\AA{} and 2.50~\AA{}, respectively. Hence, in the considered heterostructures, the graphene sheet adapts to the in-plane lattice parameter of h-BN, as observed locally in layered samples~\cite{wood+14natp,kim+17nl}. The out-of-plane lattice parameter \textit{c} is strongly dependent on the stacking arrangement, as indicated in Fig.~\ref{fig:hetero-unit cell}(b). In real samples, such relationship between stacking and interlayer distance is observed in terms of corrugations in the out-of-plane direction due to the formation of Moir$\acute{\mathrm{e}}$ patterns~\cite{argentero+17nl,wall+15ap}. Periodic vdWh, such as those considered in this work, may enhance the interlayer stability and thus reduce corrugations~\cite{haigh+12nm}.

In order to understand how the vertical atomic arrangement impacts the interlayer distance and, in turn, the stability of the vdWh, we plot in Fig.~\ref{fig:stability}(a) the in- and out-of-plane projections of the electronic charge density in the C\rotatebox{15}{$g$}B configuration, as an example. The formation energy of the considered stacking is computed according to Eq.~\ref{eq:stab}.  For reference, we also compute the formation energies of bilayer h-BN as well as of graphene interacting with either inequivalent h-BN layer in the heterostructures, as illustrated in Fig.~\ref{fig:stability}(b) and (d), left panels. All these calculations are performed in the same unit cell.

The formation energies computed in this work allow us to rank the configurations according to their relative stability. We note that we do not include contributions from phonons to the formation energies. A previous theoretical study~\cite{slotman+14adp} on similar graphene/h-BN vdWh reports that the phonon bands of the pristine constituents are basically preserved in the heterostructure, reflecting the weak vdW interactions. Therefore, a very small contribution to the formation energy is expected from zero-point vibrations, as discussed also for other vdW bound systems~\cite{milko+12prb}. Moreover, considering different stackings, it has been shown that the phonon spectra of the unfavorable configurations deviate more from those of the constituents than in the stable stackings, making them farther unstable~\cite{slotman+14adp}. Thus, the sequence of the considered stackings with respect to their relative stability, shown here, would be unaffected when including phonon contributions.
 
As shown in the bottom panel of Fig.~\ref{fig:stability}, we find that the more favorable configurations are the ones having the smallest \textit{c} parameters. The three most favorable arrangements are the C\rotatebox{15}{$g$}B, B\rotatebox{15}{$g$}B', and A\rotatebox{15}{$g$}B stackings, all with formation energies of about -0.43~eV/cell, followed by C'\rotatebox{15}{$g$}B, A'gB, and BgB with formation energies of about -0.41~eV/cell. The vertical distance between two h-BN layers is largest when the N atoms lie on top of each other. This is expected, due to the enhanced electrostatic and Pauli repulsion between the overlapping electron clouds localized around the nitrogen atoms [see Fig.~\ref{fig:stability}(a) and (b)]~\cite{hod+12jctc,leven+16jctc}. When N and B are vertically aligned, the electrostatic repulsion becomes weaker, minimizing the interlayer distance, as seen in the CgB stacking [see Fig.~\ref{fig:stability}(b)]. This characteristic was reported also for bulk h-BN~\cite{agg+18prb}. Concerning the interlayer distance between graphene and h-BN, the repulsion between N and C atoms, and thus their distance, is maximal when they are vertically aligned, giving rise to the most unfavorable stacking [see Fig.~\ref{fig:stability}(a) and (c)] as seen, for example, in the AgA configuration. The most favorable arrangement between graphene and h-BN is obtained when the B atoms lie on top of the C atoms, due to the weak repulsion forces between them [see Fig.~\ref{fig:stability}(a) and (c)]. For this reason, the stability of the BgB configuration is enhanced by $\sim$60~meV/cell (i.e., $\sim$10~meV per atom) with respect to that of the AgA stacking, even though in both configurations the h-BN layers have the same unfavorable AA stacking [see Fig.~\ref{fig:stability}(bottom panel) and (b)]. 
\begin{figure}[h!]
 \begin{center}
\includegraphics[width=.49\textwidth]{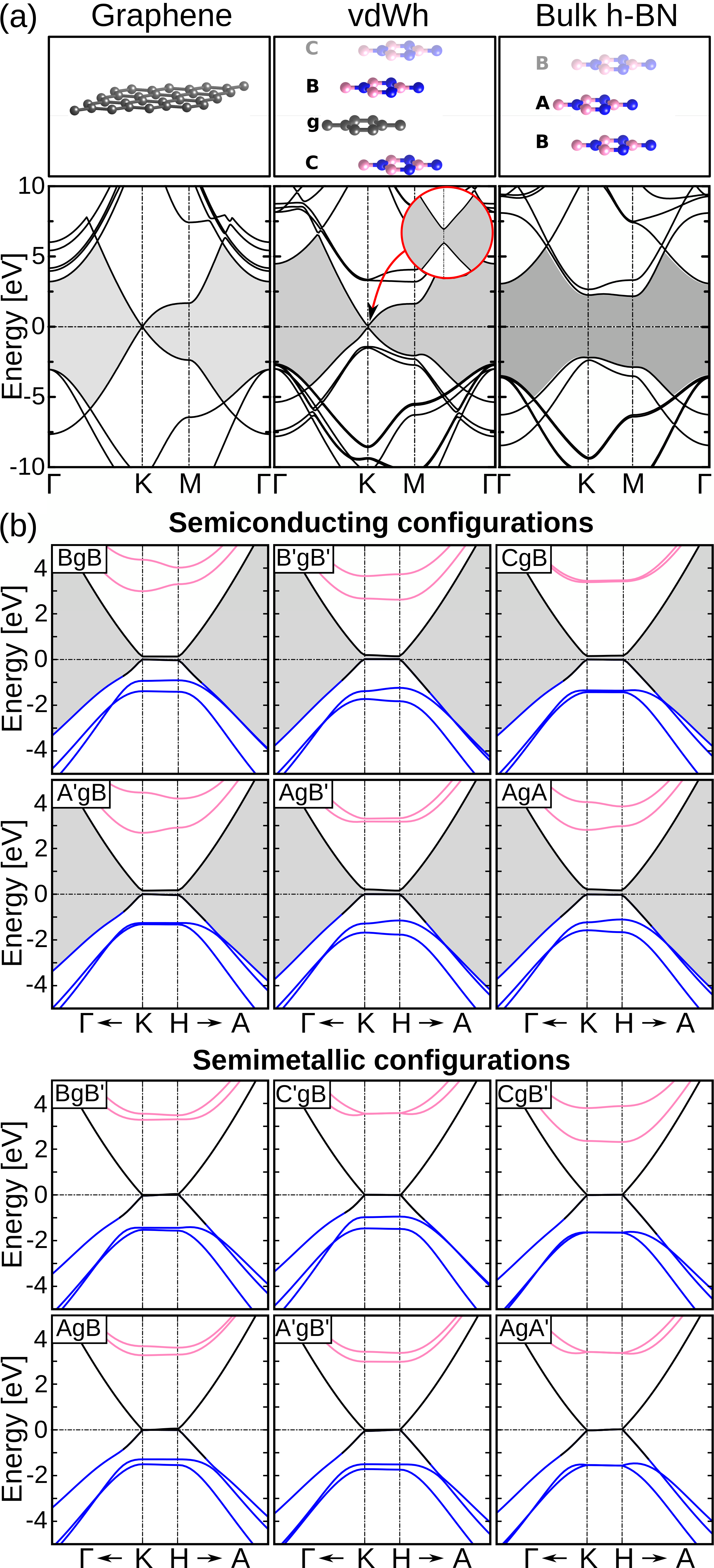}%
\caption{(a) PBE band structure of graphene, graphene/h-BN vdWh in the CgB configuration, and bulk h-BN with AB stacking. (b) PBE band structure of all considered structures. The band character is indicated by the atomic color code: pink for B, blue for N, and gray for C. The VBM is set to zero.}
\label{fig:hetero-vs-consti-str-band}
 \end{center}
\end{figure}

\section*{Electronic properties}
We now proceed with the analysis of the PBE band structure of the most stable CgB heterostructure [see Fig.~\ref{fig:hetero-vs-consti-str-band}(a)] along with those of pristine graphene and bulk h-BN in the AB arrangement. At a first glance, the band structure of this vdWh appears as a superposition of those of its constituents. This result is not surprising, considering that the covalent bonds within the layers are not altered in the assembly~\cite{agg+17jpcl}. A careful inspection, however, reveals new features arising from the weak interlayer interactions. In particular, the interaction between h-BN and graphene gives rise to a small band gap ($\sim$150~meV from PBE) at the K point (see also Refs.~\onlinecite{quhe+12npgam,agg+17jpcl})
~\cite{Note1}. The valence band maximum and conduction band minimum (VBM/CBM) are now derived from inequivalent C atoms at the K point. The energy difference between the h-BN-derived VB-1 and CB+1 states at the K point is increased by 0.27~eV in the heterostructure compared to bulk h-BN. 

In Fig.~\ref{fig:hetero-vs-consti-str-band}(b) we plot the PBE band structures of the twelve considered heterostructures in the vicinity of the (possible) gap. Interestingly, depending on the stacking arrangement, six of them are semiconducting (middle panel) and six are semi-metallic (bottom panel). The intercalation of graphene and h-BN layers breaks the lattice symmetry when inequivalent C atoms have different neighbors in the out-of-plane direction. This is the case in the AgA, CgB, and B'gB', as well as in the AgB', A'gB, and BgB configurations which are, in fact, semiconducting. On the other hand, when the two inequivalent C atoms have analogous chemical environment, as in the AgA', CgB', and C'gB configurations, as well as in the BgB' A'gB', and AgB stackings, the valence and conduction bands in graphene remain degenerate at the K point, thus preventing the opening of the gap. The resulting vdWh are therefore semi-metallic. We note in passing that the presence of flat bands in the vicinity of the Fermi energy, as observed in all the heterostructures, is a favorable condition for ferromagnetism, as reported in recent findings in the literature~\cite{repellin+20prl,chetterjee+20prb,chen+20n,liu+20npjcm}.

The band gaps of the semiconducting configurations are reported in Table~\ref{tab:hetero-stability}, where their dependence on the stacking arrangement is evident. The largest gap ($\sim$170 meV) pertains to the AgA stacking. In this case, the two inequivalent C atoms are subject to two different electrostatic potentials generated by B and N atoms, respectively, with partial charges of opposite signs [see Fig.~\ref{fig:hetero-unit cell}(b)]. As a consequence, the upshift (downshift) of the CB (VB) at the K point is maximized with respect to pristine graphene. However, this arrangement is found to be energetically unfavorable [see Fig.~\ref{fig:stability}(bottom panel)]. The smallest band gap ($\sim$130 meV) is found in the BgB stacking. In this case, only the CB band is shifted up, owing to the weak interaction between the C atom and the B atoms vertically aligned on top of it. In the CgB configuration, the layers are arranged in such a way that the interlayer repulsion is minimal, giving rise to the most favorable stacking. Moreover, due to the symmetry breaking of its sublattices, a band gap of about 150~meV is opened in graphene. A comparable value is found in the A'gB configuration, where the formation energy is reduced by about 20~meV/cell compared to that of the CgB stacking. 

Regarding the band character, in all configurations the VB and CB are dominated by C $\pi$/$\pi^*$ states, while the VB-2/VB-1 (CB+1/CB+2) by contributions from N (B) atoms [see Fig.~\ref{fig:hetero-vs-consti-str-band}(b)]. We notice, though, that the VB has N-like character away from the path K-H. We recall, that the splitting of the h-BN-derived bands along K-H can be controlled by layer stacking also in bulk h-BN~\cite{agg+18prb}. The N-like VB-2/VB-1 and the B-like CB+1/CB+2 are split along K-H when atoms of the same species are aligned on top of each other, while otherwise they are degenerate. In bulk h-BN, in presence of inversion symmetry, the wave-functions corresponding to the degenerate bands along the path K-H are distributed over both h-BN layers, except for the AB stacking. However, in all these stackings, the wave-functions at the H point are localized on one h-BN layer only~\cite{agg+18prb}. In contrast, the lack of inversion symmetry in most of the heterostructures considered here makes the wave-functions of the degenerate bands localized on only one h-BN layer along the whole K-H path. The wave-functions of the non-degenerate bands are distributed over both h-BN layers~\cite{agg+17jpcl}.

\begin{figure}[h!]
 \begin{center}
\includegraphics[width=.49\textwidth]{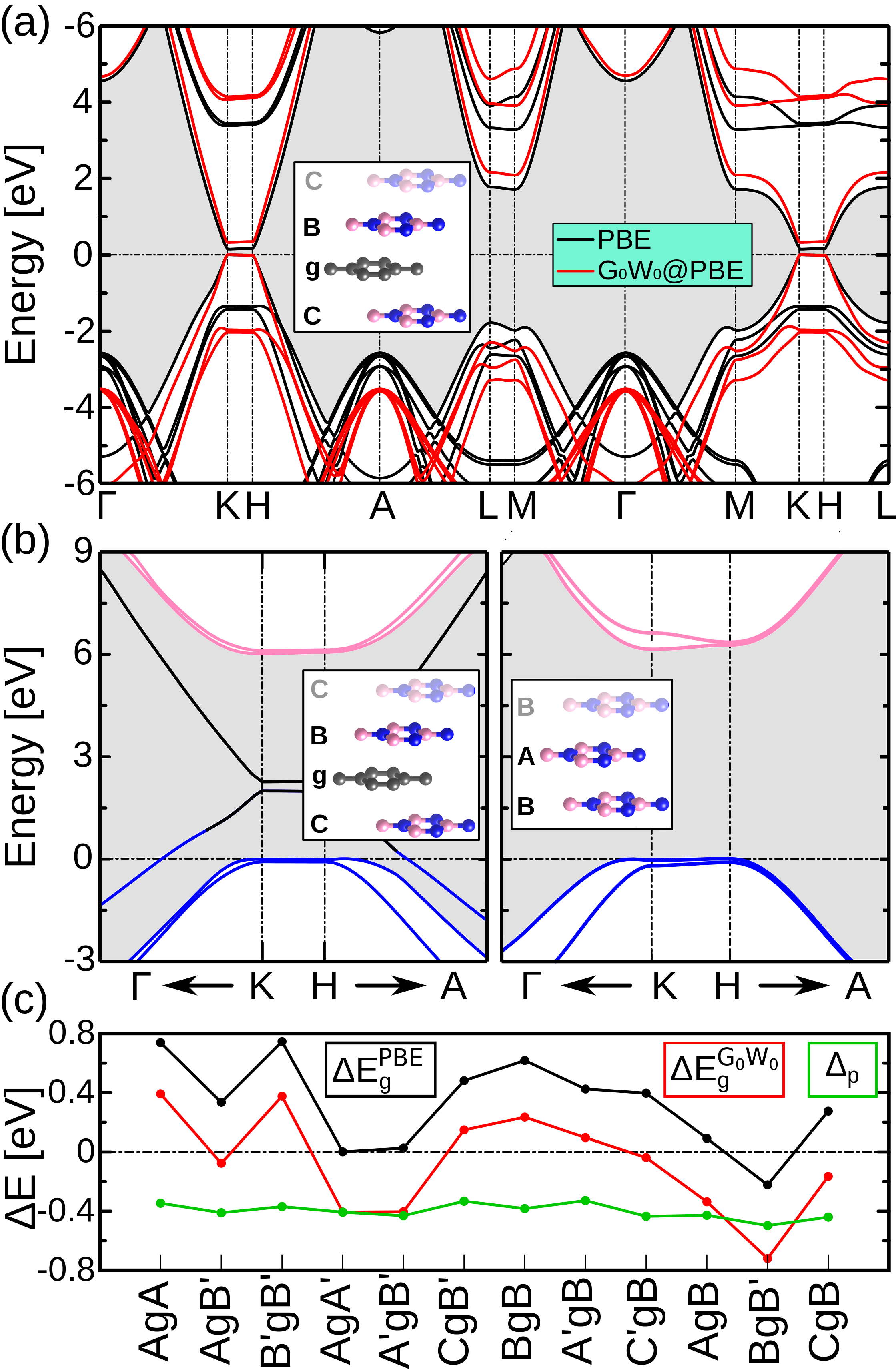}%
\caption{(a) PBE (black) and QP (red) band structure of the CgB heterostructure along the full path in the BZ, shown in Fig.~\ref{fig:hetero-unit cell}(a). The VBM is set to zero. (b) QP band structure, in the vicinity of the path K-H (left) compared with that of bulk h-BN in the AB stacking (right). The shaded area indicates the gap between the h-BN-derived bands. The band character is indicated by the atomic color code: pink for B, blue for N, and gray for C. (c) Difference between the h-BN-derived gap in the vdWh and the gap in bulk h-BN, $\Delta E$, computed from PBE (black) and $G_{0}W_{0}$ (red). $\Delta_{\mathrm{p}}$ (green) represents the renormalization of the h-BN-derived gap in the vdWh due to dynamical screening effects.}
\label{fig:hetero-vs-h-BN}
 \end{center}
\end{figure}

The electronic properties discussed so far are based on DFT results. For a quantitative description of the band structure, we need to go beyond, and apply the $G_{0}W_{0}$ approximation. In Fig.~\ref{fig:hetero-vs-h-BN}(a), we plot the QP band structure of the CgB configuration. From the comparison with the PBE result, we find that the QP correction varies from band to band. The $G_{0}W_{0}$ gap of 260~meV is approximately 100~meV larger than the PBE value (see Table~\ref{tab:hetero-stability}). However, the energy difference between the h-BN-derived bands (VB-1/CB+1 and VB-2/CB+2) increases by about 1.30~eV along the path K-H. The pronounced self-energy effect on the h-BN bands is in accordance with the insulating character of this material, characterized by localized p-orbitals at the frontier. Conversely, the delocalized $\pi/\pi^{*}$ bands in semi-metallic graphene are well captured by semi-local PBE.

\begin{table}[h]
\centering
\caption{Formation energies and electronic band gaps, $E_g$, of the considered graphene/h-BN vdWh computed with PBE and $G_{0}W_{0}$@PBE. The energy difference between the bands originating from h-BN, i.e. (VB-1/CB+1) in the vdWh at the K point, are termed ''h-BN-derived gap''.}
\vspace{0.2cm}
 \begin{tabular}{c|ccccccccccccccccccccc}
\hline
\hline
\multirow{2}*{Stacking} 
 & \multicolumn{1}{c}{E$_{form}$} && \multicolumn{3}{c}{$E_g$ [eV]} && \multicolumn{4}{c}{h-BN-derived gap [eV]}&        \\
  \cline{2-2}
 \cline{4-6}
\cline{8-11}
 &[eV/cell]&& PBE&& $G_{0}W_{0}$& &  &PBE& & $G_{0}W_{0}$ \\
\hline 
C g B   & -0.433 & &  0.15  & &0.26 & &  & 4.73  & &6.02 \\
 B g B'  & -0.428 & &  0.00  & & 0.00 & &   &4.72  & &6.01 \\
 A g B   & -0.426 & &  0.00  & & 0.00 & &   &4.55  & &5.85 \\
 C' g B   & -0.416 & &  0.00  & & 0.00 & &   &4.52  & &5.77 \\
A' g B   & -0.417 & & 0.15  & &0.27 & &   &3.94  & &5.16  \\
 B g B   & -0.411 & &  0.13 & &0.25 & &   &3.92  & &5.16   \\
 C g B'  & -0.395 & &  0.00  & & 0.00 & &   &4.00  & &5.21 \\
 A' g B'  & -0.399 & &  0.00  & & 0.00 & &   &4.49  & &5.77 \\
A g A'  & -0.389 & &  0.00  & & 0.00 & &   &4.95  & &6.31 \\
 B' g B'  & -0.357 & & 0.13  & & 0.26 & &   &4.05  & &5.29  \\
 A g B'  & -0.355 & & 0.15  & & 0.29 & &   &4.46  & &5.76  \\
 A g A  & -0.345 & & 0.17  & & 0.32 & &   &4.04  & &5.30  \\
\hline
\hline
\end{tabular}
\label{tab:hetero-stability}
\end{table}

To understand the impact of intercalated graphene on the intrinsic electronic properties of h-BN, we analyze the energy difference between h-BN-derived bands (VB-1/CB+1) in the vdWh (hereafter \textit{h-BN-derived gap}) with respect to the fundamental band gap of bulk h-BN in the corresponding stacking arrangement (see Table~\ref{tab:hetero-stability}). In Fig.~\ref{fig:hetero-vs-h-BN}(b), we compare the corresponding bands for the CgB stacking with those of bulk h-BN in the AB arrangement. At the K point, the PBE and QP gaps of bulk h-BN are 4.46~eV and 6.18~eV, respectively~\cite{agg+18prb}. By introducing a graphene sheet between the h-BN layers, the h-BN-derived gap increases by 0.27~eV, resulting in 4.73~eV. We emphasize that at the DFT level, all changes originate from electrostatic effects, including structural relaxations. Inclusion of the $G_{0}W_{0}$ self-energy leads to a decrease of the QP gap in the heterostructure to 6.02~eV compared to its counterpart in bulk h-BN ($\Delta \mathrm{E_{g}^{{G_{0}W_{0}}}}=6.02-6.18 ~\mathrm{eV}=-0.16~\mathrm{eV}$). This decrease, caused by the presence of the semi-metallic graphene sheet~\cite{thygesen+172d,neat+06prl,hues+13prb,pusc+12prb}, is due to dynamical correlation effects which are completely missed by DFT, also when adopting hybrid xc-functionals. To evaluate the magnitude of the renormalization ($\Delta_{\mathrm{p}}$) of the h-BN-derived gap due to latter effect, we subtract the changes caused by electrostatic and structural relaxation ($\Delta \mathrm{E_{g}^{\mathrm{PBE}}}$), that are already accounted for in the $\Delta \mathrm{E_{g}^{{G_{0}W_{0}}}}$ term ($\Delta_{\mathrm{p}}=-0.16-0.27=-0.43~\mathrm{eV}$). A similar renormalization value is found at the H point in the BZ. It was shown that this effect not only has a large impact on the band gap of two-dimensional  materials in the presence of metallic substrates~\cite{thygesen+172d} but yields to non-negligible contributions even in weakly polarizable systems such as interfaces of h-BN and carbon-fluoride monolayers~\cite{fu+16jpcc}. 

Considering all configurations, we find that the magnitude of the renormalization, $\Delta_{\mathrm{p}}$, is independent of the stacking arrangement [see Fig.~\ref{fig:hetero-vs-h-BN}(c)]. Differences are due to varying interlayer distances. On the other hand, we find that both $\Delta \mathrm{E_{g}^{\mathrm{PBE}}}$ and $\Delta \mathrm{E_{g}^{{G_{0}W_{0}}}}$ are strongly affected by layer stacking, due to the differences in the electrostatic interactions, which are included in both DFT and $G_{0}W_{0}$ calculations [see Fig.~\ref{fig:hetero-vs-h-BN}(c)]. The smallest values of $\Delta \mathrm{E_{g}^{\mathrm{PBE}}}$ are found in the AgA', A'gB', and AgB configurations, as well as in the CgB and BgB' alignments, where the h-BN layers have either AA' or AB stacking. The latter are the most favorable arrangements in bulk h-BN~\cite{agg+18prb}, since the repulsive interlayer forces are minimized [see Fig.~\ref{fig:stability}(b)]. Consequently, the h-BN-derived QP gap is found to be lower than in bulk h-BN in the corresponding stacking [see Fig. 4(c)]. In the AgA and B'gB' configurations, the band-gap renormalization is compensated by electrostatic effects ($\Delta \mathrm{E_{g}^{\mathrm{PBE}}}$) such that the resulting h-BN-derived QP gap is larger than in bulk h-BN~\cite{agg+18prb}. This interplay between stacking, stability, and band gap renormalization may be used as a tool to control and tailor the electronic properties of combined layered systems.

\section*{Optical excitations}
\begin{figure}[h!]
 \begin{center}
\includegraphics[width=.49\textwidth]{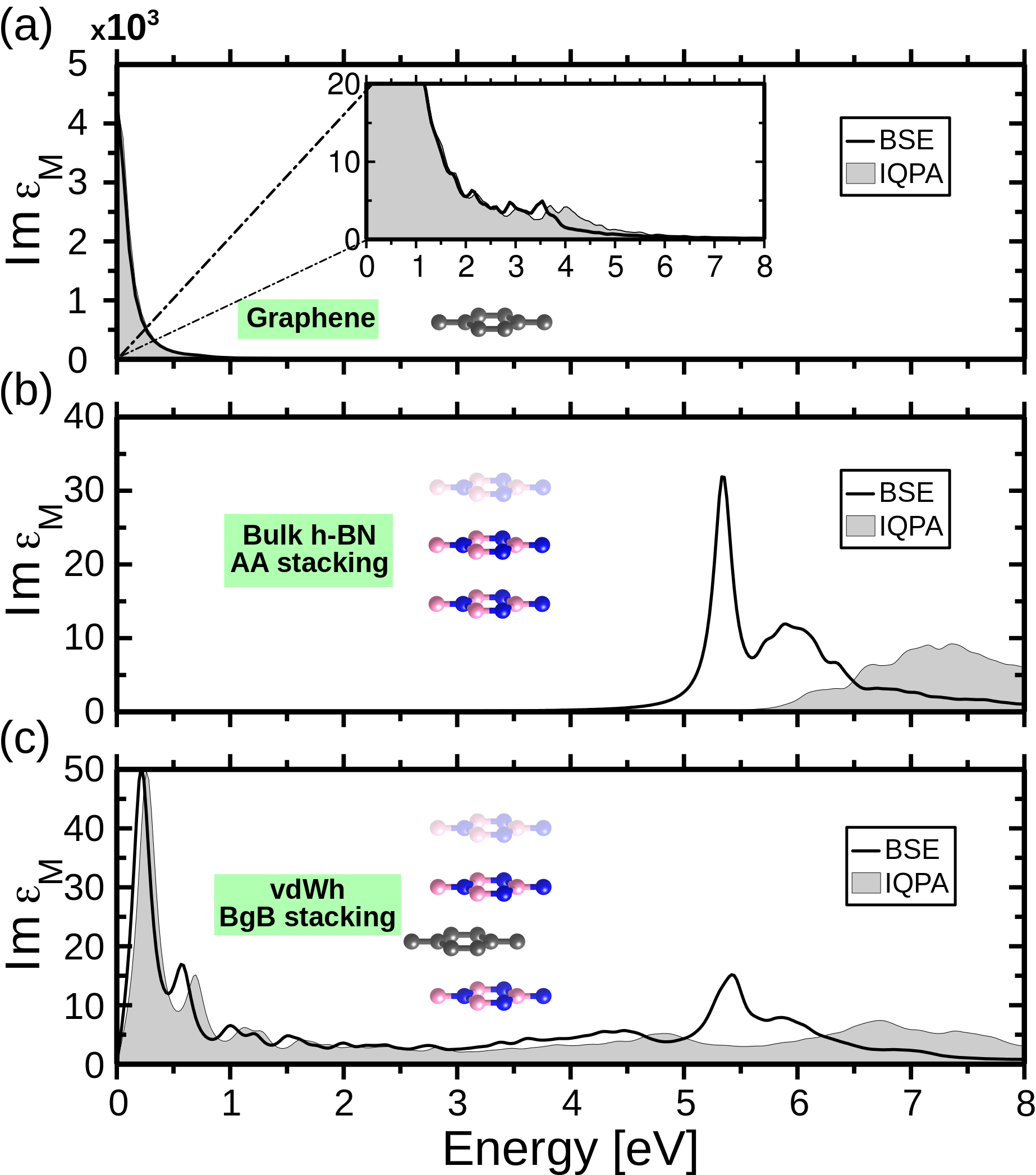}%
\caption{In-plane component of the imaginary part of the macroscopic dielectric function of (a) graphene, (b) bulk h-BN in the AA stacking, and (c) vdWh in the BgB arrangement, computed including (BSE, solid line) and neglecting (IQPA, shaded area) excitonic effects. A Lorentzian broadening of 0.1~eV is applied to all spectra to mimic the excitation lifetime.}
\label{fig:hetero-spectra-vs-constituent}
 \end{center}
\end{figure}

We conclude our analysis by investigating the optical properties of the six semiconducting graphene/h-BN heterostructures (see also Ref.~\onlinecite{agg+17jpcl}). As an exemplary case we consider in the following the BgB configuration. In Fig.~\ref{fig:hetero-spectra-vs-constituent}, we compare the optical spectra of (a) graphene, (b) bulk h-BN, and (c) the vdWh, computed with (BSE) and without (IQPA) excitonic effects. The heterostructure absorbs light over a broad frequency range, with maxima in the infrared (IR) region, below 1~eV, and in the UV range, between 5~eV and 6~eV. At visible frequencies the absorption is relatively low. These features can be easily traced back in the spectra of the constituents. Obviously, the peak in the near-IR region originates from electronic transitions within the graphene layer, which give rise to the zero-energy resonance in Fig.~\ref{fig:hetero-spectra-vs-constituent}(a)~\cite{Note2}. In the BgB heterostructure the maximum is centered at about 200~meV, due to the finite band gap of this system~\cite{Note3}. Likewise, the pronounced features in the near-UV region originate from excitons in h-BN as evident from Fig.~\ref{fig:hetero-spectra-vs-constituent}(b)~\cite{Note4}. 
 
\begin{figure*}
 \begin{center}
\includegraphics[width=.98\textwidth]{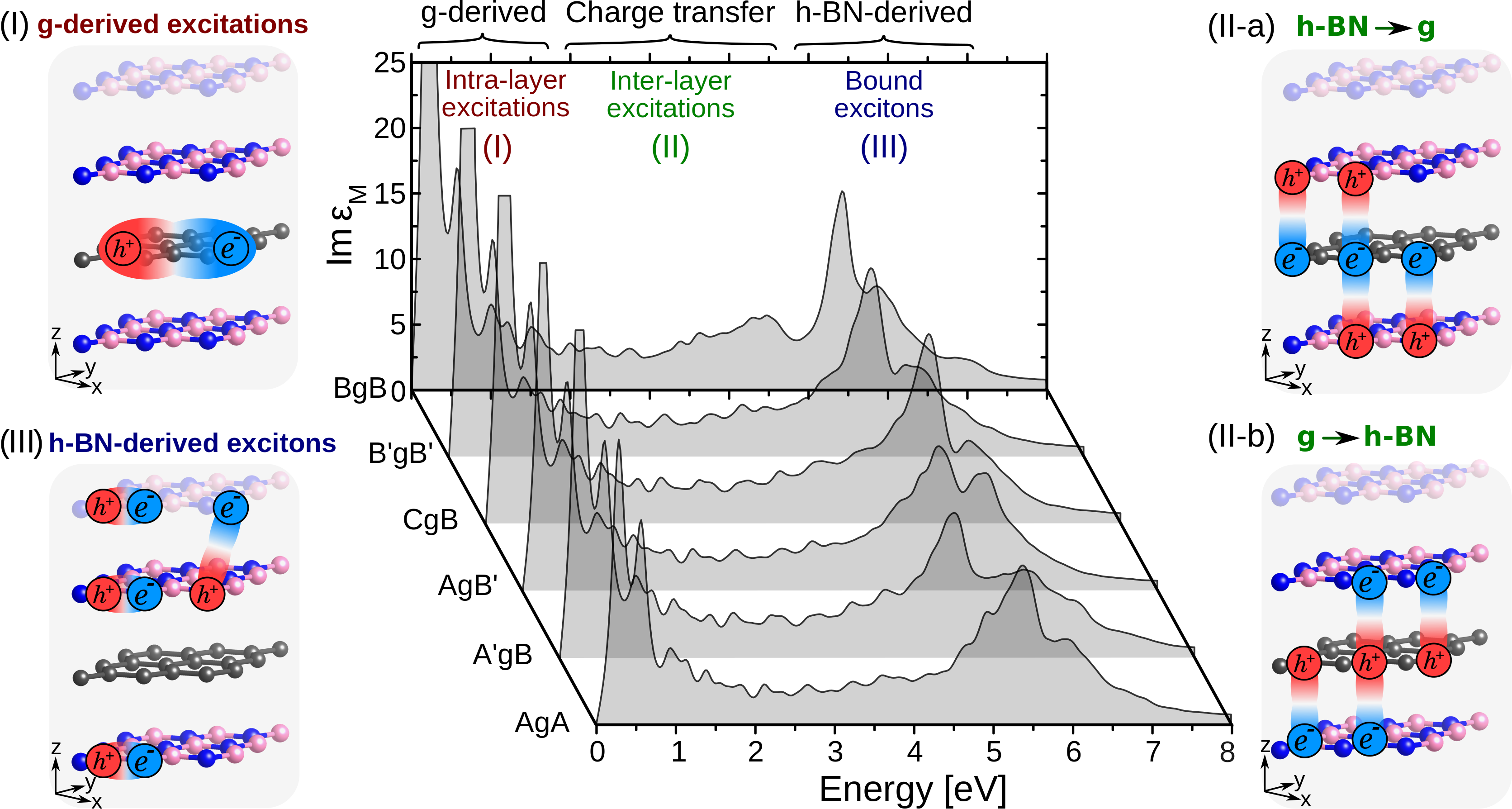}%
\caption{Central panel: In-plane component of the imaginary part of the macroscopic dielectric function of all considered semiconducting graphene/h-BN vdW heterostructures, computed from the BSE. A Lorentzian broadening of 0.1~eV is applied. Side panels: Schematic representation of the spatial distribution of the e-h pairs that dominate the regions (I, II, III) in the spectra. C, N, and B atoms are depicted in gray, blue, and pink color, respectively.}
\label{fig:spectra+excitons}
 \end{center}
\end{figure*}

To address the impact of layer stacking, we plot in Fig.~\ref{fig:spectra+excitons} the optical spectra of all semiconducting configurations. The sharp graphene-derived peak in the near-IR region appears in all arrangements. Negligible differences in the energy and intensity of these maxima reflect the variations of the QP gap (see Table~\ref{tab:hetero-stability}). The binding energy of the lowest-energy exciton is found to be 35~meV in the BgB, B'gB', CgB, and A'gB stackings.  As expected, the e-h pair is very delocalized within the graphene layer~\cite{agg+17jpcl}, as visualized in Fig.~\ref{fig:spectra+excitons} (labeled I). In the AgB' and AgA stackings, which feature the largest gap (see Table~\ref{tab:hetero-stability}), the exciton binding energy increases to 45~meV and 60~meV, respectively, reflecting the reduction of the electronic screening.

In the UV region around 5.5~eV, the sharp h-BN-derived excitonic peak is strongly affected by the  stacking (see Fig.~\ref{fig:spectra+excitons}). Depending to the specific arrangement of the h-BN sheets, the VB-2/VB-1 and CB+1/CB+2 are either degenerate or split in the gap region [see Fig.~\ref{fig:hetero-vs-consti-str-band}(b) and Table~\ref{tab:hetero-stability}] as already discussed for bulk h-BN~\cite{agg+18prb}. In the heterostructures, this feature is additionally affected by the alteration of the h-BN-derived bands due to the presence of graphene. For example, in the case of the BgB vdWh, the excitonic peak in the UV region is blue-shifted by $\sim$0.2~eV with respect to its counterpart in bulk h-BN (see Fig.~\ref{fig:hetero-spectra-vs-constituent}). This value is the result of electrostatic interactions and dynamical-screening effects [see $\Delta \mathrm{E_{g}^{{G_{0}W_{0}}}}$ in Fig.~\ref{fig:hetero-vs-h-BN}(c)]. In the CgB stacking, this excitonic peak is, instead, red-shifted by 0.16~eV with respect to bulk h-BN [see Fig.~\ref{fig:hetero-vs-h-BN}(c)]. 

The analysis of the e-h pair distribution in bulk h-BN, reported in Ref.~\cite{agg+18prb}, holds also in these heterostructures. Bound excitons in h-BN can be very localized within the layers and also exhibit charge-transfer character between h-BN layers [see (III) in Fig.~\ref{fig:spectra+excitons}]. These characteristics reflect the distribution of the atomic states involved in the VB-2/VB-1 and CB+1/CB+2 along the K-H path in the BZ~\cite{agg+17jpcl}. 

The ability to absorb light in the visible and near-UV band is a new characteristic of these vdWh that are not present in their constituents~\cite{agg+17jpcl}. The optical excitations in this region originate from transitions between bands arising from different building blocks. Consequently, the resulting excitons have charge-transfer character and binding energies of the order of 100~meV, consistent with their delocalization across two or three layers~\cite{agg+17jpcl}. These e-h pairs can be easily dissociated compared to the strongly bound excitons in bulk h-BN. 

In the visible and near-UV regions, excitons with different character are found, i.e., specific to the one of the pristine materials or mixed. In Fig.~\ref{fig:spectra+excitons}, we sketch selected interlayer excitons with characteristic spatial distribution. Excitations named II-a are formed by transitions from h-BN-derived band (VB-1 or VB-2) to the graphene-derived CB along the K-H path. This means that in all stackings the electron distribution of excitation II-a is spread over the graphene sheet. In the BgB, B'gB', AgB', and AgA arrangements the hole is distributed within both h-BN layers with the same probability (see Fig.~\ref{fig:spectra+excitons}), because the involved h-BN-derived bands (VB-1/VB-2) are split, and their corresponding wave-function is spread over the N atoms of both h-BN layers in the unit cell. On the other hand, in the CgB and A'gB stackings, the hole is located within one specific h-BN layer (see Fig.~\ref{fig:spectra+excitons}) because the h-BN-derived bands (VB-1/VB-2) involved are almost degenerate and their wave-functions are localized on the N atoms of one specific h-BN layer in the unit cell [see Fig.~\ref{fig:hetero-vs-consti-str-band}(b)]. 

Charge-transfer excitations (II-b) stem from transitions from graphene-derived (VB) to h-BN-derived band (CB+1 or CB+2). Consequently, the hole is delocalized within graphene, regardless of the stacking, while the distribution of the associated electron is tuned to reside on one or the other h-BN layer or both, depending on their relative alignment in the unit cell. Specifically, in the BgB, B'gB', A'gB, and AgA stackings the electron component of the exciton is located on both h-BN layers (see Fig.~\ref{fig:spectra+excitons}). The involved h-BN-derived bands (CB+1/CB+2) are non-degenerate along the K-H path and, hence, the corresponding wave-functions are distributed on the B atoms of both h-BN layers in the unit cell [see Fig.~\ref{fig:hetero-vs-consti-str-band}(b)]. However, in the CgB and AgB' stacking, the electron is located on one h-BN layer only, reflecting the character of the (CB+1/CB+2) electronic bands. 

\section*{Summary and Conclusions}
We have presented a first-principles many-body study of the structural, electronic, and optical properties of periodic graphene/h-BN van der Waals heterostructures. By varying the stacking arrangement, we found that the most stable configurations are characterized by the shortest out-of-plane lattice parameters, where the electrostatic and Pauli repulsions between the layers are minimal. When C and B atoms are aligned on top of each other, intercalation of graphene enhances the stability of h-BN also in configurations that are predicted to be unstable (or metastable) in the bulk structure. While the electronic bands of the constituents are basically preserved in the heterostructures, the weak interlayer interactions with the h-BN layers generates a QP band gap in graphene of the order of 300~meV. The size of the gap can be tuned depending on how the B and N atoms are aligned on top of the inequivalent C atoms. An important finding of this work is that not all graphene/h-BN heterostructures are semiconducting. In total, six out of the twelve considered configurations are semi-metallic. The polarization effects due to the intercalation of the semi-metallic graphene decreases the h-BN-derived gap by about 0.4~eV in the vdWh, regardless of the stacking. However, owing to the electrostatic interactions with graphene, the specific value of the gap depends on the arrangement of the layers. The optical spectra show that the semiconducting heterostructures absorb light over a broad frequency range, from near-IR up to the UV. At the absorption onset, transitions between graphene-derived bands are dominant. Bound excitons originating from the h-BN layers appear in the UV region. Different from their constituents, the heterostructures absorb light also at the visible and near-UV frequencies, i.e. between 1.6~eV and 4~eV. In this range, charge-transfer excitations between graphene and h-BN layers appear. The spatial distribution of the corresponding electron-hole pairs can be tuned by the stacking arrangement. As our results clearly demonstrate that the electronic structure of the constituents is basically preserved, we expect that e-h pairs in related graphene/h-BN combinations, either with different number of layers or different patterns will exhibit the same characteristics~\cite{quhe+12npgam,yan+12prb,kaloni+12jmc}.

Commenting on {\it real} materials, the flat and commensurate stackings considered in this work may appear locally in actual samples~\cite{argentero+17nl}.  As such, our results provide valuable indications to rationalize the electronic and optical properties of such microscopic regions, which can crucially affect the response of the entire system. Especially in the emerging field of Moir\'e crystals that are characterized by complex interlayer arrangements and layer rotations, our findings are useful to understand fundamental structure-property relations at the state-of-the-art level of \textit{ab initio} many-body theory.
While these methods may not be feasible to model superlattices of real cell sizes due to their high computational costs, they can provide key ingredients to develop model Hamiltonians and semi-empirical approaches that can tackle these challenges~\cite{fulvio+16prb,fulvio+2d18,sponza+19prl,sponza+18prb}.

Input and output files can be downloaded free of charge from the NOMAD Repository~\cite{drax-sche19jpm} at the following link: http://dx.doi.org/10.17172/NOMAD/2018.12.10-1.
\subsection*{Acknowledgment} 
Work supported by the Algerian Ministry of High Education and Scientific Research under the PNE programme. Partial funding by the German Research Foundation (DFG), - Project number 182087777 - SFB 951, is appreciated.

%
\end{document}